\documentclass[a4paper,english,reprint, aps]{revtex4-1}
\usepackage[T1]{fontenc}
\usepackage[latin9]{inputenc}
\usepackage{amsmath}
\usepackage{amssymb}



\usepackage{babel}
\begin{document}

\title{Exact solutions of a two-dimensional Kemmer oscillator in the gravitational
field of cosmic string}

\author{Nadjette. Messai}

\email{nadjette.messai@gmail.com}

\affiliation{Laboratoire de Physique Appliquée et Théorique, \\
 Université de Tébessa, 12000, W. Tébessa, Algeria.}

\author{Abdelmalek. Boumali}

\email{boumali.abdelmalek@gmail.com}

\affiliation{Laboratoire de Physique Appliquée et Théorique, \\
 Université de Tébessa, 12000, W. Tébessa, Algeria.}
\begin{abstract}
The two dimensional Kemmer oscillator under the influence of the gravitational
field produced by a topology such as the cosmic string spacetime and
in the presence of a uniform magnetic field as well as without magnetic
field are investigated. The eigensolutions of our problem have been
found by using the generalized parametric Nikiforov\textendash Uvarov
(NU) method, and the influence of the cosmic string space-time on
the energy spectrum has been analyzed. We show that the dependence
of the energy levels of the quantum system with the angular deficit
$\alpha$, which characterizes the global structure of the metric
in the cosmic string spacetime, breaks the degeneracy of these levels.
\end{abstract}
\maketitle

\section{introduction}

In relativistic quantum mechanics, the exact solutions of the wave
equation are very important for the understanding of the physics that
can be brought by such solutions. They are valuable tools in determining
the radiative contributions to the energy. The quantum mechanics of
charged, massive, and spin-1 particles in an external field has been
studied in many different situations using different techniques \citep{1,2,3,4,5}.
These works have especially investigated the solutions of the equation
in a magnetic field. These techniques are very complex in their determination
of the eigensolutions of such particles. The relativistic wave equation
for a massive spin-1 particle was initially derived by Kemmer in 1939.
The Kemmer equation is a Dirac-type equation, which involves matrices
obeying a different scheme of commutation rules \citep{6,7}. 

The analysis of gravitational interactions with a quantum mechanical
system has recently attracted attention in particle physics and has
been an active field of research. The general way to understand the
interaction between relativistic quantum mechanical particles and
gravity is to solve the general relativistic form of their wave equations.
These equations may be considered insignificant at the atomic scale,
where gravitational effects are weak, but the physics governing these
particles plays an important role in astrophysics and cosmology, in
which gravitational effects play a dominant role. In addition, studying
single-particle states is important to constructing a unified theory
of gravitation and quantum mechanics (see \citep{8} and references
therein) .

The Dirac oscillator was for the first time studied by Itô and Carriere.
On the other side, Moshinsky and Szczepaniak were the first who introduced
an interesting term in the Dirac equation. More specifically, they
suggested to substitute in the free Dirac equation the momentum operator
$\vec{p}$ like $\vec{p}-im\omega\beta\vec{r}$. They could obtain
a system in which the positive energy states have a spectrum similar
to the one of the non-relativistic harmonic oscillator. Recently,
this interaction has particularly got more interest \citep{9}. It
is reviewed, because of the interest in the many different domain
in physics . Furthermore, the interaction of this oscillator with
a gravitational field produced by topological defects, has become
a well-investigated topic \citep{10,11,12,13,14,15,16,17,18,19}). 

The aim of the present study is to solve the Kemmer oscillator in
a background produced by topological defects, such as cosmic strings
and magnetic cosmic strings. Also, our contribution redresses the
lack of existing literature concerning the physical properties of
charged, massive scalar particles of spin-1 interacting with gravitational
fields due to topological defects. 

This paper is organized as follows: in Sec. II, we present a review
of the the solutions of the DO in the cosmic string background. Sec.
III is devoted to solve ths case of the Kemmer oscillator in cosmic
string space-time using NU method. Finally, Sec. IV, will be a conclusion.

\section{Eigen solutions of the two-dimensional Dirac oscillator }

\subsection{The solutions without a magnetic field}

In this section we review the solutions of a two-dimensional Dirac
oscillator in the cosmic string background. The metric describe the
cosmic string is given by:

\begin{equation}
ds^{2}=-dt^{2}+d\rho^{2}+\alpha^{2}\rho^{2}d\phi^{2}+dz^{2},\label{eq:1}
\end{equation}
where $-\infty<\left(t,z\right)<+\infty$, $0<\rho\leq\infty$ and
$0\leq\phi\leq2\pi$. The parameter $\alpha$ is the deficit angle
associated with conical geometry obeying $\alpha=1-4\eta$, and $\eta$
is the linear mass density of the string in natural unite $\hbar=c=1$.

The Dirac equation in the arbitrary curved spacetime is written by\citep{11}
\begin{equation}
\left[i\gamma^{\mu}\left(x\right)\left(\partial_{\mu}-\Gamma_{\mu}\right)-m\right]\psi_{D}=0,\label{eq:2}
\end{equation}
where $m$ is the mass of particles, $\Gamma_{\mu}\left(x\right)$
are the spinor affine connections and $\gamma^{\mu}$ are the generalized
Dirac matrices satisfying the anticommutation relations
\begin{equation}
\left\{ \gamma^{\mu},\gamma^{\nu}\right\} =2g^{\mu\nu},\label{eq:3}
\end{equation}
and defined in terms of a set of tetrad fields by
\begin{equation}
\gamma^{\mu}\left(x\right)=e_{a}^{\mu}\left(x\right)\gamma^{a},\label{eq:4}
\end{equation}
where $e_{a}^{\mu}$ satisfies the relation $e_{a}^{\mu}\left(x\right)e_{b}^{\nu}\left(x\right)\eta^{ab}=g^{\mu\nu}$,
and $\left(\mu,\nu\right)=\left(0,1,2,3\right)$ are tensor indices,
$\left(a,b\right)=\left(0,1,2,3\right)$ are tetrad indices and $\gamma^{a}$
are the standard flat spacetime Dirac matrices \citep{12,13,14}.
The tensor $\eta^{ab}=\mbox{diag}\left(-1,1,1,1\right)$ is the Minkowski
tensor.

For the two-dimensional case, we choose the Dirac matrices $\gamma^{a}$
in terms of Pauli matrices as \citep{20,21,22}
\begin{equation}
\gamma^{a}=\left(\sigma^{3},i\sigma^{1},is\sigma^{2}\right),\label{eq:5}
\end{equation}
with parameter $s$ takes the values $\pm1$ ($+1$ for spin up and
$-1$ for spin down) \citep{23}.

The spinorial connection is given by\citep{22} 
\begin{equation}
\Gamma_{\mu}\left(x\right)=\frac{1}{8}\omega_{\mu ab}\left[\gamma^{a},\gamma^{b}\right].\label{eq:6}
\end{equation}
These components can be obtained by solving the Maurer-Cartan structure
equations in the absence of torsions: $d\hat{e}^{a}+\omega_{b}^{a}\hat{e}^{b}=0$,
where $\omega_{b}^{a}=\omega_{\mu\,b}^{a}\left(x\right)dx^{\mu}$,
and $\omega_{\mu\,b}^{a}\left(x\right)$ is called connection 1-form
\citep{24}. 

According to Eq. (\ref{eq:1}), we choose the tetrads for the line
elements being
\begin{equation}
\hat{e}^{0}=dt;\,\hat{e}=d\rho,\,,\hat{e}^{2}=\rho d\varphi,\,\,\hat{e}^{3}=dz.\label{eq:7}
\end{equation}
By solving the Maurer-Cartan structure equations, we obtain that 
\begin{equation}
\gamma^{\mu}\left(x\right)\Gamma_{\mu}\left(x\right)=-\frac{1}{2\rho}\gamma^{1}.\label{eq:8}
\end{equation}
Now, to include the Dirac oscillator term $im\omega\beta\rho$ into
Eq. (\ref{eq:2}), we proceed with the following substitution in the
radial momentum component $\partial_{1}\rightarrow\partial_{1}+m\omega\rho$.
Hence Eq. (\ref{eq:2}) is transformed into
\begin{align}
\left\{ i\gamma^{0}\partial_{0}+i\gamma^{1}\left(\partial_{1}+m\omega\rho\beta+\frac{1}{2\rho}\right)\right\} \psi_{D}\nonumber \\
+\left(i\frac{\gamma^{2}}{\alpha\rho}\partial_{2}-m\right)\psi_{D}=0.\label{eq:9}
\end{align}
In order to solve the Eq. (\ref{eq:9}), we adopt the following Ansatz
\begin{equation}
\psi_{D}=e^{-iEt}e^{i\left(l+\frac{1}{2}\right)\phi}\begin{pmatrix}\phi\left(\rho\right)\\
\chi\left(\rho\right)
\end{pmatrix}.\label{eq:10}
\end{equation}
By substituting (\ref{eq:10}) into (\ref{eq:9}), we get the following
system of equations
\begin{align}
\left(E-m\right)\phi,\nonumber \\
-\left\{ \left(\partial_{\rho}+\frac{1}{2\rho}-m\omega\rho\right)-is\left(\frac{l+\frac{1}{2}}{\alpha\rho}\right)\right\} \chi=0,\label{eq:11}
\end{align}
\begin{align}
\left(E+m\right)\chi\nonumber \\
+\left\{ \left(\partial_{\rho}+\frac{1}{2\rho}+m\omega\rho\right)+is\left(\frac{l+\frac{1}{2}}{\alpha\rho}\right)\right\} \phi=0.\label{eq:12}
\end{align}
After a simple algebraic calculation, we have
\begin{equation}
\left[\partial_{\rho}^{2}+\frac{1}{\rho}\partial_{\rho}-\left\{ m^{2}\omega^{2}\rho^{2}+\frac{\eta_{\pm}^{2}}{\rho^{2}}-\gamma_{\mp}\right\} \right]\begin{pmatrix}\phi\left(\rho\right)\\
\chi\left(\rho\right)
\end{pmatrix}=0,\label{eq:13}
\end{equation}
with
\begin{equation}
\eta_{\pm}=\lambda\pm\frac{1}{2},\,\left(\lambda=s\frac{l+\frac{1}{2}}{\alpha}\right),\label{eq:14}
\end{equation}
and where 
\begin{equation}
\gamma_{\mp}=E^{2}-m^{2}+2m\omega\left(\lambda\mp\frac{1}{2}\right),\label{eq:15}
\end{equation}
By making a change of variables $\varrho=m\omega\rho^{2}$ , Eq. (\ref{eq:13})
transforms into
\begin{equation}
\left\{ \frac{\partial^{2}}{\partial\varrho^{2}}+\frac{1}{\varrho}\frac{\partial}{\partial\varrho}-\left(\frac{\eta_{\pm}^{2}}{4\varrho^{2}}+\frac{1}{4}-\frac{\gamma_{\mp}}{4m\omega\varrho}\right)\right\} \begin{pmatrix}\phi\\
\chi
\end{pmatrix}\left(\varrho\right)=0.\label{eq:16}
\end{equation}
Now, in order to solve the last equation, we use the well-know method
based on the Nikiforov-Uvarof method (NU)\citep{25,26}. Thus, by
comparing with the equations from (\ref{eq:79}) to (\ref{eq:89}),
the following expressions are obtained
\begin{equation}
c_{1}=1,c_{2}=0,c_{3}=0,\label{eq:17}
\end{equation}
\begin{equation}
\xi_{1}=\frac{1}{4},\xi_{2}=\frac{\gamma_{-}}{4m\omega},\xi_{3}=\frac{\eta_{+}^{2}}{4},\label{eq:18}
\end{equation}
\begin{equation}
c_{4}=c_{5}=0,c_{6}=\frac{1}{4},c_{7}=-\frac{\gamma_{-}}{4m\omega},\label{eq:19}
\end{equation}

\begin{equation}
c_{8}=\frac{\eta_{+}^{2}}{4},c_{9}=\frac{1}{4},c_{10}=1+\begin{vmatrix}\eta_{+}\end{vmatrix},\label{eq:20}
\end{equation}

\begin{equation}
c_{11}=1,c_{12}=\frac{\begin{vmatrix}\eta_{+}\end{vmatrix}}{2},c_{13}=\frac{1}{2},\label{eq:21}
\end{equation}
\begin{equation}
\frac{2n+1}{2}-\frac{\gamma_{-}}{4m\omega}+\frac{\begin{vmatrix}\eta_{+}\end{vmatrix}}{2}=0,\label{eq:22}
\end{equation}
From these equations, the form of energy levels is
\begin{equation}
E^{2}=m^{2}+4m\omega\left(n+\frac{\begin{vmatrix}\eta\pm\end{vmatrix}}{2}-\frac{\eta\mp}{2}+\frac{1}{2}\right),\label{eq:23}
\end{equation}
with $n=\left(0,1,\cdots\right)$, $l=\left(0,\pm1,\pm2,\cdots\right)$,
and $s=\pm1$. The eigenfunction is given in terms of hypergeometric
confluent function as 

\begin{align}
\begin{bmatrix}\chi\left(\rho\right)\\
\phi\left(\rho\right)
\end{bmatrix} & =\left(m\omega\right)^{\frac{\begin{vmatrix}\eta_{\pm}\end{vmatrix}}{2}}\rho^{\begin{vmatrix}\eta_{\pm}\end{vmatrix}}e^{-\frac{m\omega\rho^{2}}{2}}\nonumber \\
 & \times F\left(-n,\begin{vmatrix}\eta_{\pm}\end{vmatrix}+1,m\omega\rho^{2}\right).\label{eq:24}
\end{align}
We note here that the eigensolutions obtained (Eqs.(\ref{eq:23},\ref{eq:24}))
are similar to the one found by \citep{11}.

\subsection{The solutions in a magnetic cosmic string background}

The two-dimensional Dirac equation in the background of a cosmic string
with magnetic field defined by its magnetic vector potential 
\begin{equation}
\overrightarrow{A}_{\phi}=i\frac{\phi_{B}}{2\pi\alpha\rho}\vec{e}_{\phi},\label{eq:24-1}
\end{equation}
 is
\begin{align}
\left\{ i\gamma^{0}\partial_{0}+i\gamma^{1}\left(\partial_{1}+m\omega\rho\beta+\frac{1}{2\rho}\right)\right\} \psi_{D}\nonumber \\
+\left(i\frac{\gamma^{2}}{\alpha\rho}\left(\partial_{2}+i\frac{e\phi_{B}}{2\pi}\right)-m\right)\psi_{D} & =0.\label{eq:25}
\end{align}
Using the same Ansatz as in Eq. (\ref{eq:10}), and project it in
Eq. (\ref{eq:25}), we obtain
\begin{equation}
\left\{ \partial_{\rho}^{2}+\frac{1}{\rho}\partial_{\rho}-\left(m^{2}\omega^{2}\rho^{2}+\frac{\eta_{\pm}^{'2}}{\rho^{2}}-\Omega_{\mp}\right)\right\} \begin{pmatrix}\phi\left(\rho\right)\\
\chi\left(\rho\right)
\end{pmatrix}=0,\label{eq:26}
\end{equation}
with the following substitutions
\begin{equation}
\eta_{\pm}^{'}=s\frac{\left[l+\frac{1}{2}\right]+\frac{e\phi_{B}}{2\pi}}{\alpha}\pm\frac{1}{2}=\lambda^{'}\pm\frac{1}{2},\label{eq:27}
\end{equation}
and 
\begin{equation}
\Omega_{\mp}=E^{2}-m^{2}+2m\omega\left(\lambda^{'}\mp\frac{1}{2}\right),\label{eq:28}
\end{equation}
Defining the following variable $\varrho=m\omega\rho^{2}$, Eq. (\ref{eq:26})
transforms into
\begin{equation}
\left\{ \frac{\partial^{2}}{\partial\varrho^{2}}+\frac{1}{\varrho}\frac{\partial}{\partial\varrho}-\left(\frac{\eta_{\pm}^{'2}}{4\varrho^{2}}+\frac{1}{4}-\frac{\Omega_{\mp}}{4m\omega\varrho}\right)\right\} \phi\left(\varrho\right)=0.\label{eq:29}
\end{equation}
By using the NU method, we have that the following eigensolutions
\begin{equation}
E^{2}=m^{2}+4m\omega\left(n+\frac{\begin{vmatrix}\eta_{\pm}^{'}\end{vmatrix}}{2}-\frac{\eta_{\mp}^{'}}{2}+\frac{1}{2}\right),\label{eq:30}
\end{equation}
with $n=0,1,2,\ldots$ and $l=0,\pm1,\pm2,\ldots$, and 
\begin{equation}
\eta_{\pm}^{'}=\lambda^{'}\pm\frac{1}{2}=\frac{s}{\alpha}\left\{ \left[l+\frac{1}{2}\right]+\frac{e\phi_{B}}{2\pi}\right\} =\eta_{\pm}^{'}=\frac{s}{\alpha}\left\{ l_{B}+\frac{1}{2}\right\} \label{eq:30.1}
\end{equation}
and where 
\begin{align}
\begin{bmatrix}\chi\left(\rho\right)\\
\phi\left(\rho\right)
\end{bmatrix} & =\left(m\omega\right)^{\frac{\begin{vmatrix}\eta_{\pm}^{'}\end{vmatrix}}{2}}\rho^{\begin{vmatrix}\eta_{\pm}^{'}\end{vmatrix}}e^{-\frac{m\omega\rho^{2}}{2}}\nonumber \\
 & \times F\left(-n,\begin{vmatrix}\eta_{\pm}^{'}\end{vmatrix}+1,m\omega\rho^{2}\right).\label{eq:31}
\end{align}
Also, these eigensolutions are similar to those found in\citep{11}.

\section{Eigen solutions of the two-dimensional Kemmer oscillator}

\subsection{The solutions without a magnetic field}

The free relativistic Kemmer equation in curved space-time is
\begin{equation}
\left(i\tilde{\beta}^{\mu}\nabla_{\mu}-M\right)\psi_{K}=0,\label{eq:32}
\end{equation}
where $M$ is the total mass of identical spin -$\frac{1}{2}$ particles
and $\tilde{\beta}$ are Kemmer matrices for the cosmic string background
given by Eq. (\ref{eq:1}). They satisfy the following commutation
relation
\begin{equation}
\tilde{\beta}^{\mu}\tilde{\beta}^{\nu}\tilde{\beta}^{\lambda}+\tilde{\beta}^{\lambda}\tilde{\beta}^{\nu}\tilde{\beta}^{\mu}=g^{\mu\nu}\tilde{\beta}^{\lambda}+g^{\lambda\nu}\tilde{\beta}^{\mu},\label{eq:33}
\end{equation}
with
\begin{equation}
\tilde{\beta}^{\mu}=\gamma^{\mu}\left(x\right)\otimes\hat{I}+\hat{I}\otimes\gamma^{\mu}\left(x\right).\label{eq:34}
\end{equation}
The $\gamma^{\mu}\left(x\right)$ are the Dirac matrices defined in
the previous section Eq. (\ref{eq:3}), $\hat{I}$ is a $4\times4$
identity matrix, and $\otimes$ indicates a direct product. The covariant
derivative in equation (\ref{eq:32}) is
\begin{equation}
\nabla_{\mu}=\partial_{\mu}-\Sigma_{\mu},\label{eq:35}
\end{equation}
where the spinorial connections can be written as \citep{22} 
\begin{equation}
\Sigma_{\mu}=\lim_{\gamma\rightarrow\sigma}\Sigma_{\mu}=\left(\Gamma_{\mu}\otimes\hat{I}+\hat{I}\otimes\Gamma_{\mu}\right),\label{eq:36}
\end{equation}
where the spinorial connection $\Gamma_{\mu}\left(x\right)$ is given
by the equation (\ref{eq:6}).

The stationary state $\psi_{K}$ of the equation (\ref{eq:32}) is
four-component wave function of the Kemmer equation, which can be
written in the form
\begin{equation}
\psi_{K}=\psi_{D}\otimes\psi_{D}=\left(\begin{array}{cccc}
\psi_{1} & \psi_{2} & \psi_{3} & \psi_{4}\end{array}\right)^{T},\label{eq:37}
\end{equation}
with $\psi_{D}$ is the solution of the Dirac equation. 

Thus, the Kemmer equation in the cosmic string background is
\begin{equation}
\left\{ i\tilde{\beta}^{0}\partial_{0}+i\tilde{\beta}^{1}\partial_{1}+i\tilde{\beta}^{2}\left(\partial_{2}-\Sigma_{2}\right)-M\right\} \psi_{K}=0.\label{eq:38}
\end{equation}
In the presence of Dirac oscillator potential, we could do the following
change: $\partial_{1}\rightarrow\partial_{1}+M\omega\rho\hat{B}$.
The operator $\mathbf{\hat{B}}$ is chosen as$\hat{B}=\gamma^{0}\otimes\gamma^{0}$
with $\hat{B}^{2}=\hat{I}$ . Hence the Kemmer equation with Dirac
oscillator interaction
\begin{equation}
\left\{ i\left(\gamma^{0}\otimes\hat{I}+\hat{I}\otimes\gamma^{0}\right)\partial_{0}+\left[\right]+\left\lceil \right\rceil \right\} \psi_{K}=0,\label{eq:39}
\end{equation}
with
\begin{equation}
\left[\right]=i\left\{ \left(\gamma^{1}\otimes\hat{I}+\hat{I}\otimes\gamma^{1}\right)\left(\partial_{1}+M\omega\rho\hat{B}\right)\right\} ,\label{eq:40}
\end{equation}
\begin{equation}
\left\lceil \right\rceil =i\left\{ \left(\gamma^{2}\otimes\hat{I}+\hat{I}\otimes\gamma^{2}\right)\left(\partial_{2}-\Sigma_{2}\right)-M\right\} .\label{eq:41}
\end{equation}
Substituting Eq. (\ref{eq:36}) into Eq. (\ref{eq:38}), we obtain
the following system of equations
\begin{align}
\left(2E-M\right)\psi_{1}-\left(\partial_{1}-M\omega\rho-\frac{is\partial_{2}}{\alpha\rho}\right)\psi_{2}\nonumber \\
-\left(\partial_{1}-M\omega\rho-\frac{is\partial_{2}}{\alpha\rho}\right)\psi_{3} & =0,\label{eq:42}
\end{align}
\begin{align}
\left(\partial_{1}+M\omega\rho+\frac{1}{\rho}+is\frac{\partial_{2}}{\alpha\rho}\right)\psi_{1}+M\psi_{2}\nonumber \\
\text{+}\left(\partial_{1}+M\omega\rho+\frac{1}{\rho}-is\frac{\partial_{2}}{\alpha\rho}\right)\psi_{4} & =0,\label{eq:43}
\end{align}
\begin{align}
\left(\partial_{1}+M\omega\rho+\frac{1}{\rho}+is\frac{\partial_{2}}{\alpha\rho}\right)\psi_{1}+M\psi_{3}\nonumber \\
\text{+}\left(\partial_{1}+M\omega\rho+\frac{1}{\rho}-is\frac{\partial_{2}}{\alpha\rho}\right)\psi_{4} & =0,\label{eq:44}
\end{align}
\begin{align}
\left(2E+M\right)\psi_{4}+\left(\partial_{1}-M\omega\rho+\frac{is\partial_{2}}{\alpha\rho}\right)\psi_{2}\nonumber \\
+\left(\partial_{1}-M\omega\rho+\frac{is\partial_{2}}{\alpha\rho}\right)\psi_{3} & =0.\label{eq:45}
\end{align}
From these equations, we get the following results
\begin{equation}
\psi_{2}=\psi_{3},\label{eq:46}
\end{equation}
 
\begin{equation}
\psi_{1}=\frac{2\left(\partial_{1}-M\omega\rho-\frac{is\partial_{2}}{\alpha\rho}\right)}{2E-M}\psi_{2},\label{eq:47}
\end{equation}
\begin{equation}
\psi_{4}=\frac{-2\left(\partial_{1}-M\omega\rho+\frac{is\partial_{2}}{\alpha\rho}\right)}{2E+M}\psi_{2}.\label{eq:48}
\end{equation}
Putting Eqs. (\ref{eq:46}), (\ref{eq:47}) and (\ref{eq:48}) into
Eq. (\ref{eq:39}), and with the following choice 
\begin{equation}
\psi_{2}=e^{iJ\phi}\chi\left(\rho\right)\label{eq:49}
\end{equation}
we get 
\begin{align}
\left(\partial_{1}^{2}+\frac{\partial_{1}}{\rho}-M^{2}\omega^{2}\rho^{2}-\frac{J^{2}}{\alpha^{2}\rho^{2}}\right)\chi\left(\rho\right)\nonumber \\
+\left(-2M\omega+4sE\omega\frac{J}{\alpha}+E^{2}-\frac{M^{2}}{4}\right)\chi\left(\rho\right) & =0.\label{eq:50}
\end{align}
Now, when we use the following transformations:
\begin{equation}
\lambda=\frac{sJ}{\alpha},\label{eq:51}
\end{equation}
\begin{equation}
\varsigma=E^{2}+4E\omega\lambda-2M\omega-\frac{M^{2}}{4},\label{eq:52}
\end{equation}
we have

\begin{equation}
\left\{ \partial_{\rho}^{2}+\frac{1}{\rho}\partial_{\rho}-\left(\frac{\lambda^{2}}{\rho^{2}}+M^{2}\omega^{2}\rho^{2}-\varsigma\right)\right\} \chi\left(\rho\right)=0.\label{eq:53}
\end{equation}
We remark that the last equation is similar to the Eq. (\ref{eq:13}).
So, by using the same method as in the case of two-dimensional Dirac
oscillator, we obtain
\begin{equation}
\left\{ \frac{\partial^{2}}{\partial\varrho^{2}}+\frac{1}{\varrho}\frac{\partial}{\partial\varrho}-\left(\frac{\lambda^{2}}{4\varrho^{2}}+\frac{1}{4}-\frac{\varsigma}{4M\omega\varrho}\right)\right\} \chi\left(\varrho\right)=0.\label{eq:54}
\end{equation}
By applying the (NU) method, we arrive at these expressions
\begin{equation}
c_{1}=1,c_{2}=0,c_{3}=0,\label{eq:55}
\end{equation}
\begin{equation}
\xi_{1}=\frac{1}{4},\xi_{2}=\frac{\varsigma}{4M\omega},\xi_{3}=\frac{\lambda^{2}}{4},\label{eq:56}
\end{equation}
\begin{equation}
c_{4}=c_{5}=0,c_{6}=\frac{1}{4},c_{7}=-\frac{\varsigma}{4M\omega},\label{eq:57}
\end{equation}
\begin{equation}
c_{8}=\frac{\lambda^{2}}{4},c_{9}=\frac{1}{4},c_{10}=1+\begin{vmatrix}\lambda\end{vmatrix},\label{eq:58}
\end{equation}
\begin{equation}
c_{11}=1,c_{12}=\frac{\begin{vmatrix}\lambda\end{vmatrix}}{2},c_{13}=\frac{1}{2},\label{eq:59}
\end{equation}
\begin{equation}
\frac{2n+1}{2}-\frac{\varsigma}{4M\omega}+\frac{\begin{vmatrix}\lambda\end{vmatrix}}{2}=0,\label{eq:60}
\end{equation}
Thus, the eigensolutions are 
\begin{align}
\psi_{2}\left(\rho\right) & =e^{-iEt}e^{iJ\phi}\left(M\omega\right)^{\frac{\begin{vmatrix}\frac{sJ}{\alpha}\end{vmatrix}}{2}}\rho^{\begin{vmatrix}\frac{Js}{\alpha}\end{vmatrix}}e^{-^{\frac{M\omega\rho^{2}}{2}}}\nonumber \\
 & \times F\left(-n,\begin{vmatrix}\frac{sJ}{\alpha}\end{vmatrix}+1,M\omega\rho^{2}\right),\label{eq:61}
\end{align}
\begin{align}
E & =2r\left(\frac{sJ}{\alpha}\right)\nonumber \\
 & \pm\sqrt{4r^{2}\frac{J}{\alpha^{2}}^{2}+2r\left(\begin{vmatrix}\frac{sJ}{\alpha}\end{vmatrix}+2n+2\right)+\frac{1}{4}}.\label{eq:62}
\end{align}
with $r=\frac{\omega}{M}$. We note here that the presence of the
parameter $\alpha$, in the spectrum of energy, breaks the degeneracy
of the energy levels. Furthermore, by taking the limit $\alpha\rightarrow1$
into Eq. (\ref{eq:62}), we reach the exact result of two-dimensional
Kemmer oscillator in Minkowskian spactime \citep{27}.

\subsection{The solutions in a magnetic cosmic string background}

By adopting the same vector potential as in the Dirac oscillator,
the Kemmer oscillator in a magnetic cosmic string spacetime obeys
\begin{align}
\left(i\tilde{\beta}^{0}\partial_{0}+i\tilde{\beta}^{1}\partial_{1}+\right)\psi_{K}\nonumber \\
+\left(i\tilde{\beta}^{2}\left(\partial_{2}-\Sigma_{2}+i\frac{e\phi_{B}}{2\pi}\right)-M\right)\psi_{K}=0.\label{eq:64}
\end{align}
By using Eq. (\ref{eq:36}), the above equation transforms into
\begin{align}
\left(2E-M\right)\psi_{1}-\left(\partial_{1}-M\omega\rho+\frac{e\phi_{B}}{2\pi\alpha\rho}-\frac{is\partial_{2}}{\alpha\rho}\right)\psi_{2}\nonumber \\
-\left(\partial_{1}-M\omega\rho+\frac{e\phi_{B}}{2\pi\alpha\rho}-\frac{is\partial_{2}}{\alpha\rho}\right)\psi_{3} & =0,\label{eq:65}
\end{align}
\begin{align}
\left(\partial_{1}+M\omega\rho+\frac{1}{\rho}-\frac{e\phi_{B}}{2\pi\alpha\rho}+is\frac{\partial_{2}}{\alpha\rho}\right)\psi_{1}+M\psi_{2}\nonumber \\
+\left(\partial_{1}+M\omega\rho+\frac{1}{\rho}+\frac{e\phi_{B}}{2\pi\alpha\rho}-is\frac{\partial_{2}}{\alpha\rho}\right)\psi_{4} & =0,\label{eq:66}
\end{align}
\begin{align}
\left(\partial_{1}+M\omega\rho+\frac{1}{\rho}-\frac{e\phi_{B}}{2\pi\alpha\rho}+is\frac{\partial_{2}}{\alpha\rho}\right)\psi_{1}+M\psi_{3}\nonumber \\
+\left(\partial_{1}+M\omega\rho+\frac{1}{\rho}+\frac{e\phi_{B}}{2\pi\alpha\rho}-is\frac{\partial_{2}}{\alpha\rho}\right)\psi_{4} & =0,\label{eq:67}
\end{align}
\begin{align}
\left(2E+M\right)\psi_{4}+\left(\partial_{1}-M\omega\rho-\frac{e\phi_{B}}{2\pi\alpha\rho}+\frac{is\partial_{2}}{\alpha\rho}\right)\psi_{2}\nonumber \\
+\left(\partial_{1}-M\omega\rho-\frac{e\phi_{B}}{2\pi\alpha\rho}+\frac{is\partial_{2}}{\alpha\rho}\right)\psi_{3} & =0.\label{eq:68}
\end{align}
From these equations, we get the following results
\begin{equation}
\psi_{2}=\psi_{3},\label{eq:69}
\end{equation}
 
\begin{equation}
\psi_{1}=\frac{2\left(\partial_{1}-M\omega\rho+\frac{e\phi_{B}}{2\pi\alpha\rho}-\frac{is\partial_{2}}{\alpha\rho}\right)}{2E-M}\psi_{2},\label{eq:70}
\end{equation}
\begin{equation}
\psi_{4}=\frac{-2\left(\partial_{1}-M\omega\rho-\frac{e\phi_{B}}{2\pi\alpha\rho}+\frac{is\partial_{2}}{\alpha\rho}\right)}{2E+M}\psi_{2}.\label{eq:71}
\end{equation}
By putting Eqs. (\ref{eq:70}) and (\ref{eq:71}) into Eq. (\ref{eq:67}),
and choosing that 
\begin{equation}
\psi_{2}=e^{iJ\phi}\chi\left(\rho\right),\label{eq:71-1}
\end{equation}
 we have
\begin{equation}
\left(\partial_{\rho}^{2}+\frac{\partial_{\rho}}{\rho}-M^{2}\omega^{2}\rho^{2}-2M\omega-\frac{\left\lfloor \right\rfloor }{\rho^{2}}+\left\langle \right\rangle \right)\chi\left(\rho\right),\label{eq:72}
\end{equation}
with
\begin{equation}
\left\lfloor \right\rfloor =\frac{J^{2}}{\alpha^{2}}+\frac{\left(\frac{se\phi_{B}}{2\pi}\right)^{2}}{\alpha^{2}}-2\frac{\left(\frac{se\phi_{B}}{2\pi}\right)sJ}{\alpha^{2}},\label{eq:73}
\end{equation}
\begin{equation}
\left\langle \right\rangle =4E\omega\left\{ \frac{sJ}{\alpha}-\frac{\left(s\frac{e\phi_{B}}{2\pi}\right)}{\alpha}\right\} +E^{2}-\frac{M^{2}}{4}.\label{eq:74.}
\end{equation}
Also, Eq. (\ref{eq:72}) can be rewritten by
\begin{equation}
\left\{ \partial_{\rho}^{2}+\frac{1}{\rho}\partial_{\rho}-\left(\frac{\mu^{2}}{\rho^{2}}+M^{2}\omega^{2}\rho^{2}-\Lambda\right)\right\} \chi\left(\rho\right)=0,\label{eq:75}
\end{equation}
with
\begin{equation}
\mu=\frac{s\left(J-\frac{e\phi_{B}}{2\pi}\right)}{\alpha},\,\Lambda=\left(E^{2}+4E\omega\mu-2M\omega-\frac{M^{2}}{4}\right).\label{eq:76}
\end{equation}
According to the above case, and by using the (NU) method, the eigensolutions
are

\begin{align}
E & =2\omega\mu\nonumber \\
 & \pm2\sqrt{\omega^{2}\mu^{2}+\left(\frac{M^{2}}{16}+M\omega\left(\frac{\left|\mu\right|}{2}+1+n\right)\right)},\label{eq:77}
\end{align}
\begin{align}
\chi\left(\rho\right) & =\left(m\omega\right)^{\frac{\begin{vmatrix}\frac{s\left(J-\frac{e\phi_{B}}{2\pi}\right)}{\alpha}\end{vmatrix}}{2}}\rho^{\begin{vmatrix}\frac{s\left(J-\frac{e\phi_{B}}{2\pi}\right)}{\alpha}\end{vmatrix}}e^{-^{\frac{m\omega\rho^{2}}{2}}}\times\nonumber \\
 & F\left(-n,\begin{vmatrix}\frac{s\left(J-\frac{e\phi_{B}}{2\pi}\right)}{\alpha}\end{vmatrix}+1,m\omega\rho^{2}\right)\label{eq:78}
\end{align}
We remark that if we put $\phi_{B}=0$ ($B=0)$ in Eq. (\ref{eq:77}),
we recovert the same results obtained in the previous section (Eq.
(\ref{eq:62})).

\section{Conclusion}

In this work, we have considered the case of a Kemmer oscillator for
vector bosons in a magnetic cosmic string space-time. The eigensolutions
are obtained by using the generalized parametric NU method. We show
that the quantum dynamics of a physical system depend on the topological
defects features of the cosmic string, and the eigensolutions possesses
an explicit dependence on the parameter $\alpha$. Furthermore, by
comparing the spectrum of energy obtained in our case with those of
the same problem in the flat spacetime \citep{27}, we can see that
the presence of the angular deficit $\alpha$, which characterizes
the global structure of the metric in the cosmic string spacetime,
breaks the degeneracies of these energies.

\appendix

\section{Review of the Nikiforov-Uvarov (NU) method}

The Nikiforov-Uvarov method is based on solving the second-order differential
equation of Schrödinger, Dirac and DKP equations by reduction to a
generalized equation of hyper-geometric type. The following equation
is a general form of the second order differential equation written
for any potential as
\begin{equation}
\left[\frac{d^{2}}{ds^{2}}+\frac{c_{1}-c_{2}s}{s\left(1-c_{3}s\right)}\frac{d}{ds}+\frac{-\zeta_{1}s^{2}+\zeta_{2}s-\zeta_{3}}{\left\{ s\left(1-c_{3}s\right)\right\} ^{2}}\right]\psi=0.\label{eq:79}
\end{equation}
According to the Nikiforov-Uvarov (NU) method, the eigenfunctions
and eigenvalues are given by
\begin{align}
\psi\left(s\right) & =s^{c_{12}}\left(1-c_{3}s\right)^{-c_{12}-\frac{c_{13}}{c_{3}}}\label{eq:80}\\
 & \times P^{\left(c_{10}-1,\frac{c11}{c_{3}}-c_{10}-1\right)}\left(1-2c_{3}s\right),\nonumber 
\end{align}
and
\begin{align}
c_{2}n-\left(2n+1\right)c_{5}+\left(2n+1\right)\left(\sqrt{c_{9}}+c_{3}\sqrt{c_{8}}\right)\nonumber \\
+n\left(n-1\right)c_{3}+c_{7}+2c_{3}c_{8}+2\sqrt{c_{8}c_{9}}=0.\label{eq:81}
\end{align}
The corresponding parameters are
\begin{equation}
c_{4}=\frac{1}{2}\left(1-c_{1}\right),\,c_{5}=\frac{1}{2}\left(c_{2}-2c_{3}\right),\label{eq:82}
\end{equation}
\begin{equation}
c_{6}=c_{5}^{2}+\zeta_{1},\,c_{7}=2c_{4}c_{5}-\zeta_{2}\label{eq:83}
\end{equation}
\begin{equation}
c_{8}=c_{4}^{2}+\zeta_{3},\,c_{9}=c_{3}c_{7}+c_{3}^{2}c_{8}+c_{6},\label{eq:84}
\end{equation}
\begin{equation}
c_{10}=c_{1}+2c_{4}+2\sqrt{c_{8}},\label{eq:85}
\end{equation}
\begin{equation}
c_{11}=c_{2}-2c_{5}+2\left(\sqrt{c_{9}}+c_{3}\sqrt{c_{8}}\right),\label{eq:86}
\end{equation}
\begin{equation}
c_{12}=c_{4}+\sqrt{c_{8}},\,c_{13}=c_{5}-\left(\sqrt{c_{9}}+c_{3}\sqrt{c_{8}}\right).\label{eq:87}
\end{equation}
In special case of $c_{3}=0$, when
\begin{equation}
\lim_{c_{3}\rightarrow0}\left(1-c_{3}s\right)^{-c_{12}-\frac{c_{13}}{c_{3}}}=e^{c_{13}s},\label{eq:88}
\end{equation}
\begin{equation}
\lim_{c_{3}\rightarrow0}\left(1-c_{3}s\right)^{-c_{12}-\frac{c_{13}}{c_{3}}}P^{\left(c_{10}-1,\frac{c11}{c_{3}}-c_{10}-1\right)}=L_{n}^{c_{10}-1}\left(c_{11}s\right),\label{eq:89}
\end{equation}
the wave function becomes
\begin{equation}
\psi\left(s\right)=s^{c_{12}}e^{c_{13}s}L_{n}^{c_{10}-1}\left(c_{11}s\right),\label{eq:90}
\end{equation}
where $L_{n}^{c_{10}-1}\left(c_{11}s\right)$ is the generalized Laguerre
polynomial.

\end{document}